\input  phyzzx
\input epsf
\overfullrule=0pt
\hsize=6.5truein
\vsize=9.0truein
\voffset=-0.1truein
\hoffset=-0.1truein

%
%

\def\IC{{\ \hbox{{\rm I}\kern-.6em\hbox{\bf C}}}}
\def\IR{{\hbox{{\rm I}\kern-.2em\hbox{\rm R}}}}
\def\IZ{{\hbox{{\rm Z}\kern-.4em\hbox{\rm Z}}}}

\def\sIR{{\hbox{{\sevenrm I}\kern-.2em\hbox{\sevenrm R}}}}

%
%
\hyphenation{Min-kow-ski}

\rightline{SU-ITP-96-12,  PUPT-1609}
\rightline{April  1996}
\rightline{hep-th/9604042}

\vfill

%
%
\title{D-branes and Fat Black Holes}

\vfill

%
%
\author{Juan Maldacena\foot{malda@puhep1.princeton.edu}}

\vfill

\address{Department of Physics \break Princeton University ,
 Princeton, NJ 08544}

\author{Leonard Susskind\foot{susskind@dormouse.stanford.edu}}

\vfill

\address{Department of Physics \break Stanford University,
 Stanford, CA 94305-4060}

\vfill

%
%

The application of D-brane methods to large black holes whose Schwarzschild
radius is larger than the compactification scale is problematic. Callan and
Maldacena have suggested that despite apparent problems of strong  
interactions
when the number of branes becomes large, the open string degrees of freedom
may remain very dilute due to the growth of the horizon area which they  
claim
grows more rapidly than the average number of open strings. Such a picture  
of
a dilute weakly coupled string system conflicts with the picture of a dense
string-soup  that saturates the bound of one string per planck area. A more
careful analysis shows that Callan and Maldacena were not fully 
consistent in their
estimates. In the form that their model was studied it can not be used to
extrapolate to large mass without being in  conflict with the Hawking
Bekenstein entropy formula. A somewhat modified model can reproduce the  
correct
entropy formula. In this ``improved model" the number of string bits  on the
horizon scales like the entropy in agreement with earlier speculations of 
Susskind.

\vfill\endpage

%
%

\REF\joe{J.~Polchinski,
 hep-th/9510017.}

\REF\stvaf{A.~Strominger and C.~Vafa,
hep-th/9601029.}

\REF\wilczek{
J. Preskill, P. Schwarz, A. Shapere, S. Trivedi and F. Wilczek,
Mod. Phys. Lett.{ A6} (1991) 2353; 
C. Holzhey and  F. Wilczek, 
 Nucl.Phys.{ B380} (1992) 447,  hep-th/9202014; 
P. Kraus and F. Wilczek,
Nucl. Phys. { B433} (1995) 403. }

\REF\vafa{
C.  Vafa, hep-th/9511088; C. Vafa, hep-th/9512078.}

\REF\holo{L.~Susskind, hep-th/9409089. }

\REF\sen{A.~Sen,   Mod. Phys. Lett.  A10 (95) 2081, hep-th/9504147.}

\REF\candj {C.~Callan, and J.~Maldacena,  Princeton University preprint
PUPT-1591,
Feb 1996, hep-th/9602043.}

\REF\spec{L.~Susskind,
Rutgers University preprint RU-93-44, August 1993, hep-th/9309145.}

\REF\garand{G.~Horowitz and A. Strominger,
hep-th/9602051.}

\REF\hms{G.~Horowitz, J.~Maldacena and A.~Strominger, hep-th/9603109.}

\REF\finn{F.~Larsen and F.~Wilczek, hep-th/9511064.}

\REF\das{S.~Das and S.~Mathur, hep-th/9601152.}

\REF\ms{J. Maldacena and A. Strominger, hep-th/9603060;
C. Johnson, R. Khuri and R. Myres, hep-th/9603061.}

%
%

%
%
\chapter{The Problem}

D-Brane methods [\joe] have made it possible to compute the degeneracy of  
the
ground states of systems which have the same quantum numbers of certain BPS
extreme
black holes [\stvaf]. The results in every case agree with the Hawking
Bekenstein
entropy, thus supporting the principle that black holes satisfy the rules of
standard quantum mechanics and statistical mechanics. Furthermore, the
development
of a large horizon for such objects confirms the idea that the  
dimensionality
of
the space of states of a region of space never exceeds the exponential of
the area
measured in planck 
units [\holo]. The D-brane methods are somewhat indirect  
and
follow a logic first outlined by Sen [\sen]. The system one is
 interested in  
is
really a gravitationally collapsed strongly interacting system. However,
using the
adiabatic invariance of the number and mass of BPS states, one can slowly  
turn
off the gravitational coupling without changing the degeneracy. The number  
of
ground states can then be computed in the limit of zero coupling where the
D-brane technology is sufficient. This method of counting is very powerful  
for
BPS states but does not really give a picture of the degrees of freedom of  
the
real gravitationally collapsed system. Thus, to understand the excited
states, their decays and issues of information transfer,
  a more direct understanding  
of
the collapsed
black hole and its degrees of freedom may be needed. 

However in  [\candj] the low lying non-BPS spectrum 
was calculated in the naive weakly coupled theory and found to 
agree with the classical result. This led to the proposal
[\candj] that 
the validity of the weakly coupled D-brane picture might extend all the way
to the
region of collapsed black holes.
Naively this seems unlikely  
because
one would expect that crowding a large number of branes together 
would lead  to a
strongly coupled mess even if the coupling constant is small.
To argue that this might  not happen  ref. [\candj] made the observation 
that gravitation causes the distribution of branes and strings to
swell up and form a macroscopic horizon with radius $R_e$. It was further
claimed that this
effect   dilutes the strings so much that they become weakly interacting.
If this
is true it is very difficult to understand the connection between these  
stringy
degrees of freedom and the dense planckian distribution of open strings that
was
derived in ref. [\spec] . We will see in what follows that the
estimates of [\candj] are not correct and that a corrected version of
the model leads to a condensate of strings consistent with ref. [\spec] .  
However,
the success of the corrected model suggests that for certain purposes the  
very
 low energy modes are  weakly interacting  with the dense soup.

We study the same five dimensional black hole solution of
type IIB string theory compactified on $T^4 \times  
S^1$
that was considered in [\candj]. For notations and
conventions we refer the reader to that reference. The system in question
consists of a collection of 1-branes and 5-branes wrapped around the
compact space. The 1-branes are wound around the $S^1$ which has radius $R$.
It will be useful to visualize the other four compact dimensions as smaller
than
$R$. However $R$ itself is to be regarded as strictly finite and fixed
throughout. The number of one and five  branes is
$Q_1$ and $Q_5$. In addition there are massless open strings carrying
Kaluza Klein
momentum $P=N/R$ in the direction of the $S^1$.
 These open strings each connect a
1-brane to a 5-brane. They can be indexed by a pair of integers $[a,  
\alpha]$
which refer to which 1-brane and 5-brane the string ends on. There are all
together
$4Q_1 Q_5$ types of such massless fermionic strings and a similar
number of bosonic strings\foot{ In fact there is also a number of (1,1) and
(5,5) strings which are  necessary to ensure that we are at a minimum of
the potential (D-flatness conditions). They are determined
by the (1,5) and (5,1) strings and are not independent degrees of
freedom (for large $Q_1,Q_5$).
}. The charges $Q_1,Q_5,N$ are
assumed to be integers and are related by
$$
c_p N = c_1 Q_1 = c_5 Q_5 = {R_e}^2
\eqn\balance
$$
where $c_p, c_1, c_5$ are constants defined in [\candj] and $R_e$ is the
Schwarzschild radius of the extremal, BPS state with those quantum numbers.

There are several domains of the parameter space that should be  
distinguished.
The first involves the limit in which the coupling constant is small enough
that
all interactions can be ignored. In this case the Schwarzschild radius is
smaller
than the string scale, $l_s =\sqrt {\alpha '}$. In this situation, reliable
state
counting can be done for both the BPS and non-BPS states. However it is only
the
BPS spectrum which can be continued to more interesting regions with any  
rigor.
In the second region the Schwarzschild radius is  much bigger than than
$l_s$ but
much smaller than $R$, the size of $S^1$. This is the region of ``black
strings".
There is some reason to think that the naive D-brane spectrum may describe  
the
spectrum of non-BPS states in this region [\garand].
 However ref. [\candj] is
interested in the much more interesting but problematic
 region in which the  
the
Schwarzschild radius is much bigger than any other scale in the problem, in
particular the compactification 
 scale $R$. This domain of parameters we will call { \it fat } black
holes.
One of the claims of [\candj] is that in  the fat  limit the strings become
infinitely
dilute on the horizon so that a weak coupling method may be sufficient to
understand not only the extreme states but also the low lying excitations.

Let us review the argument of  [\candj] for the microscopic entropy of the
BPS state. The momentum $P$ along the $X^5$ is a sum over the modes of the
massless open strings, each of which carries an integer momentum in units of
$1/R$.
$$
P={1\over R}{\sum_i{in_i}} = {N\over R}
\eqn\momentum
$$
where $n_i$ is the number of open string quanta in the $ith$ mode. The  
counting
is done with the aid of a partition function
$$
Z=\sum d(N) q^N = \left[\prod_{n=1,\infty}  
{{1+q^n}\over{1-q^n}}\right]^{4Q_1
Q_5}
\eqn\part
$$
In this expression $d(N)$ is the degeneracy of the BPS state with charges
$(Q_1,Q_5,N)$. For large $N$ ($N \gg Q_1 Q_5$) we can safely 
argue that  $d(N)$ is of order $\exp(2\pi \sqrt  
{Q_1Q_5
N})$. However,  we shall now see 
that this is incorrect in the limit in which the  
three
charges grow in fixed proportion. For simplicity we will take all three  
charges
to be equal to $N$ and calculate the asymptotic value of $d(N)$. We use the
usual trick of isolating the power series coefficients of $Z$ by contour
integration
$$
d(N) \approx \int dq {Z(q)\over q^{N+1}}
\eqn\cauchy
$$
The contour of integration surrounds the origin. A straight forward  saddle
point method can be used to estimate $d(N)$. Define $Z=\exp Q_1 Q_5 U(q)  
=\exp
N^2 U(q)$. The integral has the form
$$
d(N) \approx \int {dq\over q} {\exp
[N^2 U(q)}-N\log(q)]
\eqn\cauch
$$
The function $U(q)$ behaves like $8q$ near $q=0$. The saddle occurs at
$q={1\over 8N}$ for $N \to \infty$ and we find
$$
d(N) \to \exp \left[N+N\log 8N \right]
\eqn\dofn
$$
Thus the entropy behaves like $N \log N$. This is much smaller than 
the Bekenstein Hawking value which scales like $N^{3/2}$.
There is a simple explanation of what is happening. Consider a 1+1
dimensional gas
of free left moving massless radiation consisting of $N^2$ 
species of bosons with  
average
energy $N/R$. An effective temperature T is introduced.   

This temperature is the effective temperature of the left movers
and should not be confused with the Hawking temperature which is
related to the effective temperature of the 
right movers [\candj ]. Assume that the entropy,
 energy and number of quanta are
extensive.
Then, on dimensional grounds, the entropy 
and number of particles will be
proportional to one another (this ignores
a mild logarithmic   infrared
divergence in the
number of quanta that we will return to). Up to numerical factors 
the energy
and entropy satisfy 
$$
E=N^2 R T^2 ={N\over R}
\eqn\energy
$$
$$
S=N^2 R T
\eqn\entropy
$$
Eliminating the temperature $T$ gives
$$
S=N^{3/2}
$$
which is consistent with the results of   [\candj]. But now calculating 
the temperature
$$
T={1\over {R N^{1/2}}}
$$
we notice that as $N$ grows  $T$ is driven to zero. At some point the  
wavelength $1/T$ of
a typical quantum exceeds $R$ and the assumption of extensivity fails. 
In fact it is straight forward to show that the characteristic thermal 
length scale $R N^{1/2}$ is nothing but the Schwarzschild radius $R_e$. 
In other words extensivity fails when the black hole becomes fat.

Before discussing how to fix the model for fat black holes we would like to
discuss two other difficulties that we will  see are related. The first
involves the density   of
open strings that forms on the horizon. These open strings are strikingly
similar to the objects discussed by one of us [\spec]. These open strings
are expected to become very dense on the horizon with $\approx1$ open
string per planck area. However   [\candj]   estimates the density of
 strings
and find that it goes to zero as the mass and charge get large. To see why,
return to eq.
\momentum .  Since the  
i's
are all positive integers it is evident that the total number of strings is
bounded by $N$. Furthermore the horizon area grows like $N^{3/2}$. Thus the
number of strings per unit area goes to zero at least as fast as  
$N^{-{1\over2}}$. One may also say that the number of strings per unit
entropy is going to zero.
This hypothetical behavior is to be contrasted with the situation 
in which extensivity prevails. In this case the number of quanta 
scales in the same way as the entropy. For example in the case of
fermions the average total number of quanta at temperature $1/\beta$ is
given by the integral
$$
n_f=R\int d\omega {1\over{e^{\beta\omega}+1}} \approx RT
\eqn\fquanta
$$
For bosons the corresponding integral is mildly divergent
$$
n_b=R\int d\omega {1\over{e^{\beta\omega}-1}} \approx RT\log (RT)
\eqn\fquanta
$$
Note that the logarithm comes from quanta with wave length longer than 
$1/T$ but these infrared quanta do not contribute significantly to 
either the energy or the entropy.

The second point involves the mass gap to the first excited 
state above the  BPS
state. We can excite the system by adding a pair of oppositely
moving open
stings in the lowest mode. Since 
the lowest mode has energy $1/ R$ this  
gives a
gap of order
$$
\delta M  \approx 1/R
\eqn\gapo
$$
 This is very strange in the limit in which the largest
dimension of the black hole is the Schwarzschild radius. A much more  
plausible
value would involve some negative power of $R_e$. It is surprising that
ref. [\candj ]  was  able to reproduce the very low temperature
thermodynamics with a theory that produces much too large a gap.
%
%
\chapter{The Fix}

It is important to understand that there is nothing wrong with 
the value of  the
BPS entropy reported in  [\candj]. There are more rigorous ways to
obtain it by using supersymmetry and analytic continuation from a region of
parameters where the D-brane model is reliable [\stvaf ].
 It is the use of the naive
version of the model in a domain of parameters where it does not apply 
which  is at fault.
In what follows we will suggest a modification of the model which
simultaneously fixes all three problems discussed in section 1. 
The model is
based on an observation of Das and Mathur [\das].   Let us begin with an
analogy from
elementary quantum mechanics. Consider a circular loop of wire of unit  
radius
whose center is at the origin of the $r,\theta$ plane. A bead of unit mass
moves
on the wire and for obvious reasons the angular momentum of the bead is
quantized
in integer multiples of
$\hbar$. The energy spectrum is given by
$$
E={l^2\over 2}
\eqn\qm
$$
for all integer $l$.
 Now consider a wire which is wrapped $n$ times around the same circle.  
Eq.
\qm\
 still gives the energy levels if we allow $l$ to be an integer multiple  
of
$1/n$. The system simulates fractional angular momentum. The real physical
system
of wire plus bead must, of course, have integer angular momentum but the  
energy
spectrum may be expressed in terms of a 
``psuedo-angular momentum" which is  
not
the true generator of spatial rotations but rather the generator of  
rotations
of
the bead relative to the physical wire.

Next let us consider a set of $Q_1$  1-branes wrapped on $S^1$, ignoring for
the
time being, the 5-branes. In a very naive way we may distinguish the
various ways
the branes interconnect. For example they may connect up so as to form one  
long
brane of total length $R'= R Q_1$. At the opposite extreme they might
form $Q_1$ disconnected loops. The spectra of open strings is different in  
each
case. For the latter case the open strings behave like $Q_1$ species of
1 dimensional particles, each  with energy spectrum given by integer
multiples of
$1/R$. In the former case they behave more like a single species of 1
dimensional
particle living on a space of length $Q_1 R$. The result  [\das] is a
 spectrum  
of
single
particle energies given by integer multiples of $1\over{Q_1 R}$  .
 In other  
words
the system simulates a spectrum of   fractional charges. For consistency the
total
charge must add up to an integer multiple of $1/R$ but it can do so by  
adding
up
fractional charges. Note that in this case, as opposed to the
bead and wire example, the branes by themselves cannot carry any 
momentum since they are invariant under boosts along directions
parallel to the branes.

Now let us return to the case of both 1 and 5 branes. By suppressing  
reference
to
the four compact directions orthogonal to $x^5$ we may think of 
the 5 branes  
as
another kind of 1 brane wrapped on $S^1$. The 5-branes
 may also be connected  
to
form a single multiply wound brane or several singly wound branes. Let us
consider the spectrum
 of (1,5) type strings (strings which connect a 1-brane  
to
a five-brane) when both the 1 and 5 branes each
form a single long brane. The 1-brane has total length $Q_1 R$ and the  
5-brane
has length $Q_5 R$. A given open string can be indexed by a pair of indecies
$[a,\alpha]$ labelling which loop of 1-brane and 5-brane it ends on. As a
simple
example choose $Q_1=2$ and $Q_5=3$. Now start with the $[1,1]$ string which
connects the first loop of 1-brane to the first loop  of 5-brane. Let us
transport this string around the $S^1$. When it comes back to the starting
point
it is a $[2,2]$ string. Transport it again and it becomes a $[1,3]$
 string.  
It
must be cycled 6 times before returning to the $[1,1]$ configuration. It
follows
that such a string has a spectrum of a single species living on a circle of
size
$6R$. More generally, if $Q_1$ and $Q_5$ are relatively prime the system
simulates a single species
 on a circle of size $Q_1 Q_5 R$. If the $Q's$ are  
not
relatively prime the situation is slightly 
more complicated but the result is the same.  For example  suppose
$Q_1 = Q_5 =Q$,  again  assume the 5 and 1-branes each form a single long  
brane,
then a string will return to its original configuration after
cycling 
around $Q$ times. This time the system simulates $Q$ species living  
on
a
circle of length $Q$. But  it is also possible to remove one loop from  
either
the 1 or 5 brane and allow it to form a separate disconnected  loop. In this
case we have a system consisting of a brane of length $QR$, one of length
$(Q-1)R$ and a short loop of length $R$. Since $Q$ and $Q-1$ are relatively
prime the open strings which connect them live on an effective brane of  
length
$Q(Q-1)R$. Thus there is always a way of hooking up branes so that the
effective length is of order $Q_1 Q_5 R$. In fact we will argue
that this type of configurations give the largest entropy, and will
therefore be dominant.

It can also be seen from the original derivation 
of the black hole entropy by 
Vafa and Strominger [\stvaf ], that the system should have low 
energy modes with energy of order $ 1/R Q_1 Q_5 $.
In this derivation  the degrees of freedom that 
carry  the momentum were described by a superconformal 
field theory on the orbifold $(T^4)^{Q_1 Q_5} /S(Q_1 Q_5)$. 
A careful analysis of this theory shows that the low 
energy modes are present.
This again corresponds to a particular
way of wrapping the branes, since the ground state degeneracy
corresponds to that of a {\it single } wound string (in the
U-dual picture) with winding and momentum $(n,m)=(Q_1,Q_5)$ 
[\vafa ].

If, on the other 
hand, the system consists of singly wound 1 and 5 branes then
there are effectively $Q_1 Q_5$ species of
open strings living on a circle  
of size $R$. This is the case analyzed in ref. [\candj] .

We can easily see that this model gives the correct value for the extremal
entropy. Let us consider the case where
$Q_1$ and
$Q_5$ are  relatively prime. As in [\candj] the open strings have 4 bosonic 
and 4 fermionic degrees of freedom and carry total
momentum $N/R$. This time the quantization length is 
$R'=Q_1Q_5R$ and the momentum is quantized in 
units of $(Q_1 Q_5 R)^{-1}$. Thus instead of being at level
$N$ the system is at level $N'=NQ_1Q_5$. In place of the original
$Q_1Q_5$
species we now have a single specie. To compute the extremal 
entropy we may introduce a partition function analogous to eq.\part\
with the appropriate replacements.
The result is 
$$
S=2\pi\sqrt{N'}=2\pi\sqrt{NQ_1Q_5}
\eqn\S
$$
 A qualitative argument
can be given by taking the $Q's$ to be of order $N$ and 
assuming extensivity. Again suppressing numerical constants 
$$
E=N/R = R' T^2 =N^2 R T^2
\eqn\ener
$$
$$
S=R' T =N^2 R T
\eqn\ent
$$
The equations are exactly the same as eqs. \energy\ and \entropy\
and the effective temperature is unchanged. However this  time the
condition for extensivity is that
$$
R'>T^{-1}
$$
which is easily satisfied. The picture is somewhat reminiscent of 
that proposed in [\finn]
although the details differ.
Next consider the energy gap.  Obviously the gap 
is given by replacing $R$  
by
$Q_1Q_5R$ in eq. \gapo
$$
\delta M = {1\over(R')} ={1\over(Q_1 Q_5 R)}
\eqn\gap
$$
Using formulae in ref. [\candj ]  one finds that the gap scales like
$$
\delta M ={{G_N}\over R_e^4}
\eqn\gapa
$$
where $G_N$ and $R_e$ are the Newton constant and Schwarzshild radius.

It is very interesting that this same
result can also be obtained from a thermodynamic argument
 [\wilczek ]. 
 The authors in [\wilczek ]  argue that thermodynamics of near extremal
black holes  will only break  down when
the temperature is so low that the specific heat is of order unity. For a  
five
dimensional black hole this happens at a temperature
$T \approx {G_N\over R_e^4}$. In order for the black hole to be able to  
radiate
at such low temperature the gap should be of order eq. \gapa. The very
long length scale
$$
R_g={{{R_e}^4}\over G_N}
\eqn\gapscale
$$ 
associated with the gap has no obvious 
analog for a Schwarzschild black hole.

For the four dimensional black holes studied in [\ms ] the same 
arguments lead to a mass gap $ \delta M = 1/R Q_2 Q_5 Q_6 $. This 
agrees with the  point at which the classical thermal approximation
breaks down, which is  $ \delta M \sim {G_N \over R_e^3 } $ in
four dimensions.

Having gotten a reasonable behavior for the mass gap we may hope 
to reproduce the correct behavior of the near-extreme non-BPS
entropy. The  arguments of [\candj] can once again
be applied to the fixed model. As in [\candj] there will be a
contribution to the near extremal entropy from right movers. If the
right movers carry momentum $N_R/R$ then by an argument identical to 
that preceding eq.\S\ we find an incremental entropy
$$
\delta S=2\pi\sqrt{N'_R}=2\pi\sqrt{N_R Q_1Q_5}
\eqn\Sright
$$
This  accounts for only a third of the contribution needed in 
order to agree with the entropy obtained from the classical 
black hole solution. However as explained in [\hms]
the classical solutions indicate that an additional contribution 
from antibranes must be present. If we naively apply U-duality which
interchanges $N$,$Q_1$ and $Q_5$ we can account for the other $2/3$ of
the incremental entropy. This argument clearly needs to be tightened.

Finally, we may also borrow the argument concerning the evaporation
process of near extreme black holes from [\candj]. Instead of $Q_1Q_5$ 
species of open strings we have only a single specie. The right moving 
temperature is again identified as the Hawking temperature and is
unchanged from [\candj].The factor $Q_1Q_5$ appearing in eq(4.6) of
[\candj] now appears as a volume factor for the gas of particles.

Our last point concerns the relation between
the D-brane picture and the arguments given
in refs.[\holo] and [\spec]. In [\spec]  arguments were 
made to show that the degrees of freedom that 
give rise to horizon entropy are open strings attached to 
the horizon. It was also argued in [\holo] that no matter how small the 
string coupling $g$ may be, when an area-density of
$1/g^2$ (in string units)
is achieved, interactions become important.
This ``saturation density" 
defines both the planck density and the 
density of open strings on the
horizon of any black hole. As we have seen, 
whenever extensivity prevails, as it does in
the ``fixed" model, the entropy and number of 
quanta scale the same way.  
Thus, apart from the very infrared quanta, 
the density of open
strings on the puffed up horizon  will always be of order one per 
planck unit of area
in agreement with [\spec].

As a consequence of the high density of string it is 
generally unlikely that weak coupling methods, including the use 
of D-branes, can be used
to  quantitatively study the interaction of a black 
hole with infalling matter. However, there may be an exception range 
of parameters which is amendable to weak coupling approximations
although we can offer no clear proof of this. The success
of the model in the very long wave length region suggests 
that the infrared quanta with wave length larger than $R_e$ but smaller than
than the very long length scale $R_g$  may  behave like goldstone bosons and
decouple from the dense  horizon soup. In this
respect we might recall an entirely different, but apparently 
analogous,  physical situation: electrons and nuclei in a metal. The low
energy thermodynamics can be fairly well 
reproduced by considering the electrons to be free, though there
are lots of charges present. In that case we have a better understanding as
to which  physical questions can be answered by regarding the electrons as
free and which require taking into account the interactions.
Hopefully, further studies of these models will provide a similar
understanding for the fat black hole case.

What is the justification for picking the configurations in which
the 1 and  5
branes each form single long branes and ignoring other configurations? The
answer lies in the entropy of each configuration. As we have seen, the
entropy of the configuration in which  the branes form a maximal number  of
short loops of size $R$ is only of order $N\log{N}$ while the entropy of the
state with maximally long connected   components
is of order $N^{3\over 2}$.  In fact,
the configurations with the maximum  
entropy are always those  in  which the branes behave  as if
they were a single  long  brane.

\ack

We are  indebted to Curt Callan, Barak Kol, Edi Halyo and Arvind Rajaraman
for   helpful discussions . The research of
J.M. was supported in part by DOE grant DE-FG02-91ER40671. 
L.S. was was supported in part by NSF grant  PHY 9219345

\refout
\end